
\input phyzzx
\hoffset=0.2truein
\voffset=0.1truein

\def\slash#1{\rlap{$#1$}/}

\def\tbar{ {${\bf \overline 3}$} }
\def\sext{ {\bf 6} }
\def\xicz{	\Xi_{c1}^0	}
\def\xicp{	\Xi_{c1}^+	}
\def\lamc{	\Lambda_c^0	}
\def\xicczs{	\Xi_{c2}^{0*}	}
\def\xiccz{	\Xi_{c2}^{0}	}

\def\xiccp{	\Xi_{c2}^{+}	}

\def\sigcz{	\Sigma_{c}^{0}	}

\def\sigcp{	\Sigma_{c}^{+}	}

\def\sigcpp{	\Sigma_{c}^{++}	}
\def\omec{	\Omega_c^0	}

\input epsf
\def\INSERTFIG#1#2#3{\epsfysize=#1in
      \hbox to\hsize{\hfil\epsffile{#2}\hfil} {\sl #3} }

\frontpagetrue

\hbox to\hsize{
  \hfil\vtop{\hbox{\strut CMU-HEP95-18} \hbox{\strut DOE-ER/40682-106}}}

\vskip .5in \centerline{\fourteenpoint The Radiative Charmed Baryon Decay
$\Xi_{c2}^{0*}
\rightarrow \Xi_{c1}^0 \gamma$}
\vskip .3in \centerline{Ming Lu\footnote{\dag}{Email address:
lu@fermi.phys.cmu.edu},
Martin J.~Savage\footnote{\ddag}{Email address: savage@thepub.phys.cmu.edu}
and James Walden\footnote{\S}{Email address: walden@fermi.phys.cmu.edu}}
\medskip
\centerline{\it Department of Physics, Carnegie Mellon University,}
\centerline{\it Pittsburgh, PA 15213, USA.}

\vfill
\vskip .2in \vfil \centerline{\twelvepoint \bf Abstract}
V-spin symmetry ($s \leftrightarrow d$ symmetry)
forbids the radiative decay $\Xi_{c2}^{0*} \rightarrow \Xi_{c1}^0 \gamma$
in the SU(3) limit.
The quark mass term breaks V-spin symmetry and the leading nonanalytic
contribution to the radiative decay amplitude is computable in
heavy baryon chiral perturbation theory.
The radiative decay branching ratio is determined by the coupling constant
$g_2$
and at leading order in chiral perturbation theory is given by
 $Br(\Xi_{c2}^{0*} \rightarrow \Xi_{c1}^0 \gamma) = 1.0\times 10^{-3} g_2^2$.
Measurement of this branching fraction will determine $|g_2|$.
\vfil

\vfill\noindent
October, 1995

\vfil\eject\pageno=1

Recently CLEO reported the discovery of $\Xi_{c2}^{0*}$ ($J^\pi = {3\over
2}^+$, sextet)
with a mass $m(\Xi_{c2}^{0*}) = 2643\pm 2$ MeV [1]. The dominant decay mode
of $\Xi_{c2}^{0*}$ is to $\Xi_{c1}$ ($J^\pi = {1\over 2}^+$, anti-triplet) and
a pion.
Radiative decay $\Xi_{c2}^{0*} \rightarrow \Xi_{c1}^0 \gamma$ is forbidden
in the SU(3) limit by V-spin symmetry (a symmetry of strong and electromagnetic
interactions under the interchange of strange and
down quarks), an SU(2) subgroup of flavor SU(3).
We point out that the leading contribution to the radiative decay amplitude of
$\Xi_{c2}^{0*}$ is nonanalytic in quark masses ($\sim {\cal O}(m_q^{1/2})$),
finite and computable in heavy baryon chiral perturbation theory.
This leading order calculation gives a radiative decay branching ratio
$Br(\Xi_{c2}^{0*} \rightarrow \Xi_{c1}^0 \gamma) = 1.0\times 10^{-3} g_2^2$,
where $g_2$ is the ${\bf 6^{(*)}6^{(*)}}\pi$ coupling in the heavy baryon
chiral Lagrangian.
Using the value of $g_2$ in the large $N$ limit of QCD (where $N$ is the number
of colors)
yields a branching ratio of $\sim {1\over 3}\%$.
Although it may be difficult to observe a branching fraction
of $10^{-2}$ at CLEO [2], the E781 experiment at Fermilab may be able to
reach
the $10^{-3}$ level [3]. This would be sufficient to
determine $|g_2|$ to an accuracy of $\sim 30\%$.
Previous work on radiative charmed baryon decays found that the decay amplitude
of  $\Xi_{c2}^{0*} \rightarrow \Xi_{c1}^0 \gamma$ is small [4].
However,  measurement of this branching ratio
will be one of the only ways to determine $|g_2|$.
Since $g_2$ enters in many heavy baryon loop calculations, its determination
is vital if we wish to go beyond tree level in the heavy baryon sector.
Furthermore, it is important to test
the large $N$ and quark model predictions for axial coupling constants such as
$g_2$.

Physics of hadrons containing a single heavy quark simplifies significantly
in the limit where the mass of the heavy quark becomes infinitely greater than
the scale of strong interactions [5]. In the heavy quark limit,
heavy hadrons can be classified according to the spins of the light
degrees of freedom, $s_\ell$. The lowest lying charm baryons contain
the $s_\ell = 0 (J^\pi = {1\over 2}^+)$ states which transform as
\tbar under flavor SU(3) ($\Lambda_c^0, \Xi_{c1}^0$ and
$\Xi_{c1}^+$), and the $s_\ell = 1 (J^\pi = {1\over 2}^+, {3\over 2}^+)$
states which transform as \sext under flavor SU(3) ($\Sigma_c^{++(*)},
\Sigma_c^{+(*)}, \Sigma_c^{0(*)}, \Xi_{c2}^{+(*)}, \Xi_{c2}^{0(*)}$ and
$\Omega_{c}^{0(*)}$) [6].
It is convenient to introduce two superfields $T_i(v)$ (which transforms as
\tbar)
and $S^{ij}(v)$ (which transforms as \sext), where $v$ (the velocity of charmed
baryons)
is conserved in the heavy quark limit. The two superfields are
$$  T_i (v) = {1 + \slash{v} \over 2} B_i ~~,  $$
$$  S_\mu^{ij} (v) = {1 \over \sqrt{3}} (\gamma_\mu + v_\mu) \gamma_5
  {1 + \slash{v} \over 2} B_{ij} + {1 + \slash{v} \over 2} B_\mu^{*ij} ~~.
\eqno (1)  $$
Here
$$  B_1 = \xicz~,~~ B_2 = -\xicp~,~~ B_3 = \lamc, ~~ \eqno (2)  $$
while
$$  B_{11} = \sigcpp~,~~ B_{12} = {1 \over \sqrt{2}} \sigcp~,~~ B_{22} =
\sigcz, $$
$$  B_{13} = {1\over \sqrt{2}} \xiccp~, ~~B_{23} = {1\over \sqrt{2}} \xiccz~,~~
  B_{33} = \omec, ~~  \eqno (3)  $$
with the corresponding $B_{ij}^{(*)}$ fields for spin-${3\over 2}$ partner of
\sext.

Interactions of heavy hadrons with soft pions and photons (of energy
$\ll \Lambda_{\chi} \sim 1$ GeV) can be described by using both heavy quark
symmetry and chiral symmetry. At lowest order in the heavy quark expansion
and chiral expansion the heavy baryon chiral Lagrangian
is given by [6]
$$  {\cal L} = {f^2\over 8} Tr(\partial_\mu \Sigma \partial^\mu \Sigma^\dagger)
+
  {\bar T}^i iv\cdot D T_i - {\bar S}_{ij}^\mu iv\cdot D S_\mu^{ij}
  + \Delta {\bar T}^i T_i   $$
$$  + ~ig_2 \epsilon_{\mu\nu\lambda\rho} {\bar S}_{ij}^\mu v^\nu
(A^\lambda)_k^i
  S^{\rho,jk} + g_3 (\epsilon_{ijk} {\bar T}^i (A^\mu)^j_l S_\mu^{kl} + h.~c.~)
  ~~, \eqno (4)  $$
where $\Delta$ is the mass difference between the \sext and \tbar states,
and $D_\mu$ is the chiral covariant derivative.
The vector and axial vector chiral fields $V_\mu$ and $A_\mu$
are formed from the pseudo-Goldstone boson fields
$$  V_\mu = {1\over 2}(\xi^\dagger \partial_\mu \xi + \xi \partial_\mu
\xi^\dagger) ~, $$
$$  A_\mu = {i\over 2}(\xi^\dagger \partial_\mu \xi - \xi \partial_\mu
\xi^\dagger) ~,
  \eqno (5)  $$
and $\xi^2 = \Sigma = \exp\left({2iM\over f}\right)$, where $M$ is the meson
octet
$$  M = \pmatrix{{1 \over \sqrt2} \pi^0 + {1 \over \sqrt6} \eta & \pi^+ & K^+
\cr
\pi^- & - {1 \over \sqrt2} \pi^0 + {1 \over \sqrt6} \eta & K^0 \cr
K^- & \bar K^0 & - \sqrt{2 \over 3} \eta \cr}~,   \eqno (6)   $$
and $f \simeq 132$ MeV is the pion decay constant.

The axial coupling constants $g_2$, $g_3$ are unknown parameters in the
effective
theory and must be determined experimentally. It has been shown that in the
large $N$
limit they are given by $g_3 = \sqrt{3\over 2} g_A, ~g_2 = -{3\over 2} g_A$
[7,8],
where $g_A$ is the nucleon axial coupling.
(Experimentally $g_A = 1.25$.)
One expects that the deviations from these $N = \infty$ relations occur at the
$1/N$
level, i.e., $\sim 30\%$ [8].
(The nonrelativistic quark model gives $g_3 = \sqrt2$, $g_2 = -2$.)
The total hadronic decay width of $\xicczs (\rightarrow \xicz\pi^0,
\xicp\pi^-)$
is given at tree level by
$$  \Gamma_0 = {g_3^2\over 8\pi} {|\vec p_\pi|^3 \over f^2} ~.
  \eqno (7) $$
The CLEO result
$\Gamma_0 < 5.5$ MeV [1] gives an upper bound $|g_3| < 1.4$~.
At present there is no experimental information on $g_2$.
It is important to determine $g_2$, not only because $g_2$ enters in many heavy
baryon loop computations, but also because this will indicate how well the
large $N$
and quark model predictions work in the heavy baryon sector.

We now turn to the radiative decays of the $J^P = {3\over 2}^+$ baryons
in the \sext of SU(3). The leading contribution to $\sext \rightarrow
{\bf \overline 3} \gamma$ is the magnetic dipole (M1) radiation with
the electric dipole (E2) component suppressed by a factor of $1/m_c$ [9].
In chiral Lagrangian the M1 transition arises from a dimension $5$ operator
$$  {\cal L} = a {e\over \Lambda_{\chi}} {\bar T}^i {\cal Q}^j_l \gamma^\mu
   \gamma_5 S^{kl,\nu} F_{\mu\nu} \epsilon_{ijk} ~~,   \eqno (8)  $$
where
$$  {\cal Q} = {1\over 2} (\xi Q \xi^\dagger + \xi^\dagger Q \xi) ~,  \eqno (9)
 $$
$Q$ is the charge matrix for the light quarks
$$  Q = \pmatrix{{2\over 3} & & \cr
                 & -{1\over 3} & \cr
                 & & -{1\over 3} \cr} ~~,  \eqno (10)  $$
and $a$ is some unknown constant of order ${\cal O}(1)$ by dimensional
analysis.
This term gives the leading contribution to the radiative decays
$\Sigma_c^{0(*)} \rightarrow \lamc \gamma$ and
$\Xi_{c2}^{+(*)} \rightarrow \xicp \gamma$.
However, it doesn't contribute to $\Xi_{c2}^{0(*)} \rightarrow \xicz \gamma$
because of the V-spin symmetry. The V-spin symmetry is broken by quark masses
($m_d \not= m_s$), and the leading contribution to
$\Xi_{c2}^{0(*)} \rightarrow \xicz \gamma$
arises from the 1-loop graph (fig.~1) by keeping the masses of $K$ and $\pi$
in the loop. This gives rise to an amplitude of order
${\cal O} \left(m_q^{1/2}\right)$.
An explicit calculation gives the partial width for
$\xicczs \rightarrow \xicz \gamma$
\footnote{\dag}{The same calculation gives the width of $\xiccz$. (Pionic
transitions
from $\Xi_{c2}$ to $\Xi_{c1}$ are kinematically forbidden.) }
$$  \Gamma_\gamma = {E_\gamma^3 \over 6\pi} \left({eg_2g_3\over
16\pi^2f^2}\right)^2
  \left[{1\over 2}E_\gamma \log{m_K^2\over m_\pi^2} -
  \left(J(m_K^2, E_\gamma) - J(m_\pi^2, E_\gamma)\right)\right]^2 ~~,  \eqno
(11)  $$
where
$$  J(m^2, E) = \int_0^1 dx \sqrt{m^2 - x^2E^2 - i\epsilon}
  \left(\pi - 2 \tan^{-1}{xE \over \sqrt{m^2 - x^2E^2 - i\epsilon}}\right)
  ~,  \eqno (12)  $$
and $E_\gamma$ is energy of the photon.
(In the limit $\Delta \rightarrow 0$, $J(m^2) \rightarrow \pi m$.)
{}From equations (11) and (12), we arrive at the branching ratio for
$\xicczs \rightarrow \xicz \gamma$
$$  \eqalign{Br&(\xicczs \rightarrow \xicz \gamma) = {\Gamma_\gamma \over
\Gamma_\pi} \cr
  &= {\alpha g_2^2 \over 3 \pi}
  {E_\gamma^3 \over |\vec p_\pi|^3} {1\over (4\pi f)^2}
  \left[{1\over 2}E_\gamma \log{m_K^2\over m_\pi^2} -
  \left(J(m_K^2, E_\gamma) - J(m_\pi^2, E_\gamma)\right)\right]^2  ~\cr
    &= 1.0\times 10^{-3} g_2^2 ~~.\cr} ~  \eqno (13)  $$
Inserting the large $N$ value for $g_2$ yields
a radiative branching ratio of $\sim {1\over 3}\%$. This is probably within
the reach of E781, and may also be seen at CLEO with some luck.

The leading counter terms which contribute to $\xicczs \rightarrow \xicz
\gamma$
have the form
$$  {\cal L} = b_1 {e \over \Lambda_{\chi}^2} {\bar T}^i {\cal Q}^j_l
   (\xi m_q \xi + \xi^\dagger m_q \xi^\dagger)^k_m \gamma^\mu \gamma_5
S^{lm,\nu}
   F_{\mu\nu} \epsilon_{ijk}   $$
$$  + ~b_2 {e \over \Lambda_{\chi}^2} {\bar T}^i {\cal Q}^j_m
   (\xi m_q \xi + \xi^\dagger m_q \xi^\dagger)^m_l \gamma^\mu \gamma_5
S^{kl,\nu}
   F_{\mu\nu} \epsilon_{ijk} ~,  \eqno (14)   $$
where $m_q$ is the light quark mass matrix
$$  m_q = \pmatrix{m_u & & \cr
                   & m_d & \cr
                   & & m_s \cr} ~~,  \eqno (15)  $$
and $b_1$, $b_2$ are unknown constants of order ${\cal O}(1)$.
These contributions are suppressed by a factor of order
${\cal O}\left(m_s^{1/2}\right)$ relative to the leading nonanalytic piece.
There are also ${\cal O}\left({1\over m_c}\right)$ SU(3)
breaking corrections to the above result.
{}From naive dimensional analysis we expect that these higher order
contributions are down
by a factor of ${\cal O}\left({m_K \over m_c}, {m_K\over \Lambda_\chi}\right)$
and therefore we expect our result to hold with an approximately $30\%$
uncertainty.

The electric quadrupole (E2) contribution to $\xicczs \rightarrow \xicz \gamma$
violates both the heavy quark spin symmetry and flavor SU(3).
(The E2 transition to decay $\Sigma_c^* \rightarrow \Lambda_c \gamma$ was
considered
in ref. [9].) Formally, the leading contribution to the
E2 amplitude comes from the same graph (fig.~1)
as the M1 amplitude but with the $\sext^*-\sext$ mass difference explicitly
retained.
A straightforward calculation gives the ratio of the E2 amplitude to the M1
amplitude
to be approximately $1\%$. (The mass splitting between the $\sext^*$
and $\sext$ states is taken
to be $\sim 64$MeV [10].) This is too small to be seen in the near future.

To conclude, we have studied the SU(3) breaking charm baryon decay
$\xicczs \rightarrow \xicz \gamma$ at leading order in chiral perturbation
theory
and found a branching ratio of
$1.0\times 10^{-3} g_2^2$, where $g_2$ is the
${\bf 6^{(*)}6^{(*)}}\pi$ axial coupling constant in the heavy baryon chiral
Lagrangian.
We estimate that the theoretical uncertainty of our result is approximately
$30\%$.
Measurement of this branching fraction may prove to be the best way
(and perhaps the only way) to determine $|g_2|$.

\bigskip

\centerline {\bf Acknowledgement}

We thank J. Yelton and J. Russ for useful discussions.
This work was supported in part by DOE under contract DE-FG02-91ER40682.
MJS acknowledges partial support from the DOE Outstanding Junior Investigator
program.

\bigskip\bigskip

\centerline {\bf References}

\item{1.} CLEO Collaboration, P.~Avery et al., CLNS 95/1352, CLEO 95-14,
\hfil\break
hep-ex/9508010.

\item{2.} J. Yelton, private communication.

\item{3.} J. Russ, private communication.

\item{4.} H.-Y. Cheng et al., Phys. Rev. {\bf D49} (1994)5857.

\item{5.} N. Isgur and M.B. Wise, Phys. Lett. {\bf B232} (1989) 113; Phys.
Lett.
{\bf B237} (1990) 527; H.~Georgi, Phys. Lett. {\bf B240} (1990) 447.

\item{6.} P. Cho, Phys. Lett. {\bf B285} (1992) 145.

\item{7.} Z. Guralnik, M. Luke and A.V.~Manohar, Nucl. Phys. {\bf B390} (1993)
474.

\item{8.} E.~Jenkins, Phys. Lett. {\bf B315} (1993) 431.

\item{9.} M.J. Savage, Phys. Lett. {\bf B345} (1995) 61.

\item{10.} M.J. Savage, Phys. Lett. {\bf B359} (1995) 189.

\bigskip\bigskip
\bigskip

\vfill\eject

\INSERTFIG{2.5}{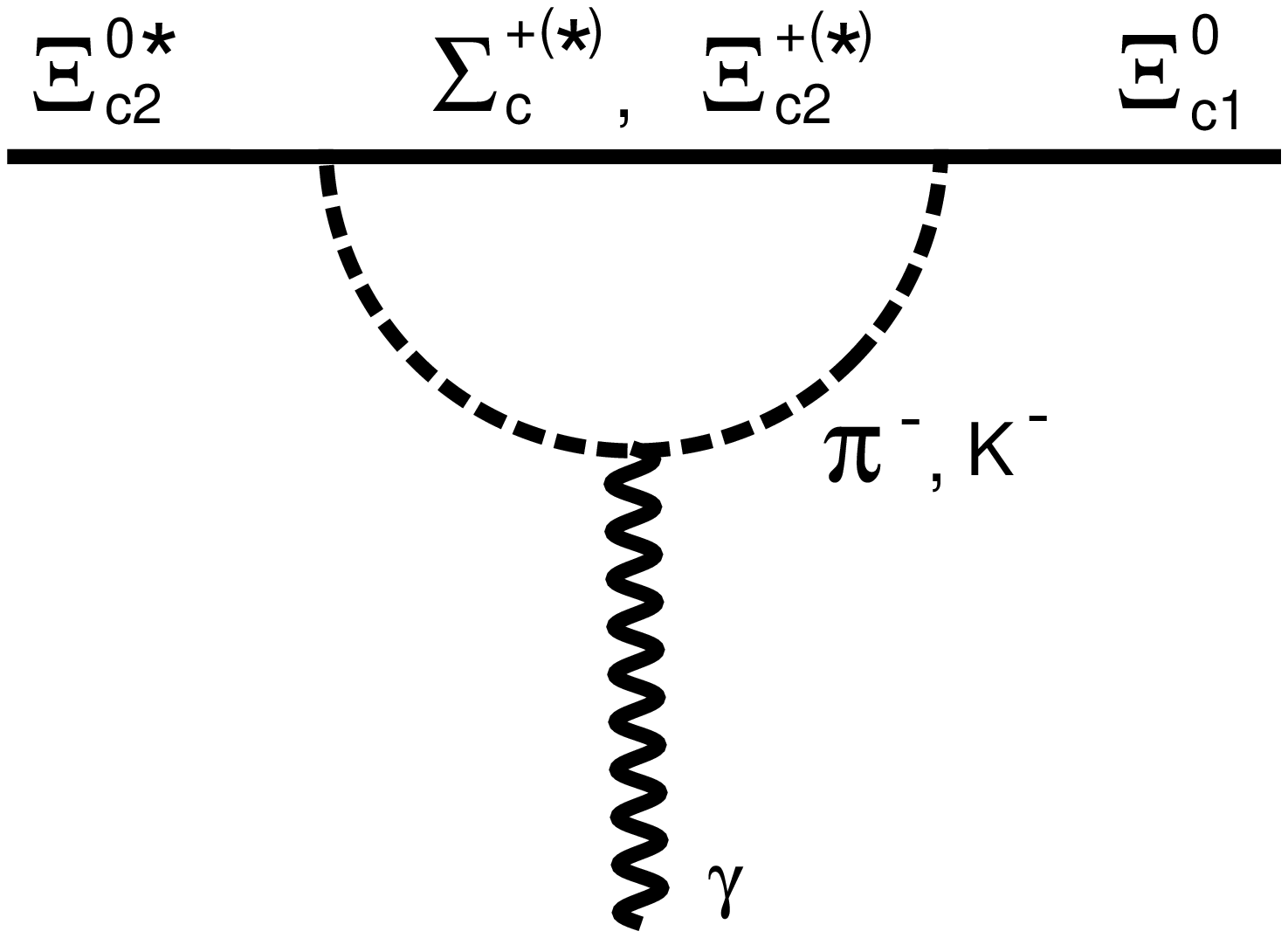}{Fig.~1: Feynman diagram contributing to $\xicczs
\rightarrow
\xicz \gamma$ at leading order in heavy baryon chiral perturbation theory.}

\vfill\eject

\end